\documentclass[apj]{emulateapj}

\usepackage{amsmath,amstext,amssymb}
\usepackage{graphicx}
\usepackage{hyperref}
\usepackage[figure,figure*]{hypcap}
\usepackage{enumerate}
\usepackage{multirow}
\usepackage{bm}
\usepackage{natbib}
\usepackage{amsfonts}
\usepackage{amssymb} 
\usepackage{bm}
\usepackage{natbib}
\usepackage{courier}
\usepackage{xcolor}

\mathchardef\mhyphen="2D

\newcommand{\del}[1]{}

\newcommand{\comm}[1]{}

\shorttitle{The CHiPS Survey: Introduction}
\shortauthors{Somboonpanyakul et al.}

\altaffiltext{\MIT}{Kavli Institute for Astrophysics and Space Research, Massachusetts Institute of Technology, 77 Massachusetts Avenue, Cambridge, MA 02139}
\altaffiltext{\Princeton}{Jadwin Hall, Princeton University, Princeton, NJ 08540, USA}
\altaffiltext{\LSST}{LSST, 950 N. Cherry Ave, Tucson, AZ 85719, USA}
\altaffiltext{\CfA}{Harvard-Smithsonian Center for Astrophysics, 60 Garden St., Cambridge MA 02138, USA}

\def\MIT{1}
\def\Princeton{2}
\def\LSST{3}
\def\CfA{4}

\begin{document}

\title{The Clusters Hiding in Plain Sight (CHiPS) survey: A first discovery of a massive nearby cluster around PKS1353-341}

\author{
Taweewat Somboonpanyakul\altaffilmark{\MIT},
Michael McDonald\altaffilmark{\MIT},
Henry W. Lin\altaffilmark{\Princeton},
Brian Stalder\altaffilmark{\LSST}, and
Antony Stark\altaffilmark{\CfA}
}

\begin{abstract}
We introduce the first result of the Clusters Hiding in Plain Sight (CHiPS) survey, which aims to discover new, nearby, and massive galaxy clusters that were incorrectly identified as isolated point sources in the \textit{ROSAT} All-Sky Survey. We present a \textit{Chandra} X-ray observation of our first newly discovered low-redshift ($z = 0.223$) galaxy cluster with a central X-ray bright point source, PKS1353-341. After removing the point source contribution to the cluster core ($L_{nuc}\sim1.8\times10^{44}$ $\rm{erg\,s^{-1}}$), we determine various properties of the cluster. The presence of a relaxed X-ray morphology, a central temperature drop, and a central cooling time around 400 Myr indicates that it is a strong cool-core cluster. The central galaxy appears to be forming stars at the rate of $6.2\pm3.6\,M_\odot\,\rm{yr^{-1}}$, corresponding to $\sim$1\% of the classical cooling prediction. The supermassive black hole in the central galaxy appear to be accreting at $\sim$0.1\% of the Eddington rate with a total power output of $\sim$5$\times10^{45}$ $\rm{erg\,s^{-1}}$, split nearly equally between radiative and mechanical power. We see weak evidence of localized excess entropy at a radius of 200 kpc which, if true, could imply a recent ($\sim$180 Myr) energetic outburst in the core that has risen buoyantly to a larger radius. Comparing the cluster's bulk properties with those of other known clusters (e.g., the total mass $M_{500}$ is $6.9\substack{+4.3\\-2.6}\times10^{14}\,M\rm{_\odot}$, and the bolometric X-ray luminosity $L_X$ is $7\times10^{44}$ $\rm{erg\,s^{-1}}$), we show that this cluster, which is massive enough that it was detected (but not confirmed) by the Planck survey, is also sufficiently luminous that it would have been identified as a cluster in the \textit{ROSAT} All-sky Survey if it did not have such a bright central point source. This discovery demonstrates the potential of the CHiPS survey to find massive nearby clusters with extreme central properties that may have been missed or misidentified by previous surveys.

\end{abstract}

\keywords{galaxies: clusters: general -- galaxies: clusters: intracluster medium -- X-rays: galaxies: clusters}

\section{Introduction}
Clusters of galaxies are the largest collapsed objects in the universe~\citep{2005Voitb}. Because of the deep gravitational potential well of clusters, supernovae and active galactic nuclei (AGN) are unable to expel gas beyond the turnaround radius, allowing the study of galaxy formation in a closed system. Simulations with radiative cooling and gravity alone are not sufficient to explain the observed properties of a brightest cluster galaxy (BCG), the most luminous galaxy in a cluster~\citep{Balogh2001,2007Bregman,2012McNamara}. The simulations tend to predict too much cool gas and too many newborn stars. This is referred to as \textit{the cooling flow problem}, and the best candidate for explaining this discrepancy is kinetic feedback by the central AGN in clusters~\citep{2006Bower,2008Bower,2006Croton}.

There are two primary modes of AGN feedback which allow the supermassive black hole (SMBH) at the center of its host galaxy to affect the final stellar mass of the galaxy. The first mode is the kinetic mode, driven by radio jets, and the second mode is the quasar mode, or radiative mode, which relates to radiation from the accretion disk~\citep[see reviews by][]{2012Fabian,2012McNamara}. The kinetic mode has been intensively studied, specifically in galaxy clusters, which require feedback to prevent overcooling~\citep{2006Rafferty,2007McNamara} via radio jets and bubbles~\citep{2011Fabian}. In contrast, the impact of radiative feedback on clusters is poorly understood~\citep{1998Silk}, due to the relative lack of central cluster galaxies in the quasar mode and a smaller region in the center in which the radiative feedback (compared to mechanical feedback) can be observed.

There are only four known examples of galaxy clusters hosting central quasars: H1821+643~\citep{2010Russell}, 3C 186~\citep{2005Siemiginowska, 2010Siemiginowska}, IRAS 09104+4109~\citep{2012OSullivan}, and the Phoenix cluster~\citep{2012McDonald}. The small number of such objects is insufficient to fully exemplify the role of radiative feedback in the evolution of galaxy clusters and their central BCGs, including the duty cycle of radiative feedback, its correlation with radiative cooling, and the distinction between the effects of type I and type II quasars on clusters~\citep{2015Kirk}. One possible way to uncover more of these objects comes from the surprise discovery of the Phoenix cluster, which, at $z = 0.6$, is the most X-ray luminous cluster known and the nest of a massive central starburst~\citep{2012McDonald,2013McDonald,2015McDonald}. While this cluster was initially discovered with the Sunyaev Zel'dovich effect~\citep{2011Williamson}, further investigation reveals that it had previously been detected by several all-sky surveys at a variety of wavelengths, but had consistently been classified as an isolated AGN because of the extremely active central galaxy and a (relative) lack of extended X-ray emission due to its distance. This leads us to wonder how many nearby ($z<0.7$) galaxy clusters with central quasars or massive starbursts are currently mislabeled in existing all-sky surveys.

In this work, we present the Clusters Hiding in Plain Sight (CHiPS) survey to look for galaxy clusters misidentified in existing surveys due to the extreme nature of their central galaxies. In this pilot study, we present a newly discovered galaxy cluster, surrounding the quasar PKS1353-341, along with new \textit{Chandra} observation of the galaxy cluster and its central AGN. By performing a detailed study of this object, we can deduce the properties of the cluster and investigate the impact a central quasar has on the intracluster medium (ICM). 

The data used in the CHiPS survey and its methodology are described in Section~\ref{sec::chips}. In Section~\ref{sec::chandra}, we present the \textit{Chandra} analysis. The results and discussion are presented in Sections~\ref{sec::result} and~\ref{sec::diss}. We assume $H_0 = 70\,\rm{km/s/Mpc}$, $\Omega_m = 0.3$ and $\Omega_{\lambda} = 0.7$. All errors are $1\sigma$ unless noted otherwise.

\section{The CHiPS Survey} \label{sec::chips}
The CHiPS survey is designed around the idea that centrally concentrated galaxy clusters at high redshift or clusters hosting extreme central galaxies (starbursts and/or QSOs) can have been misidentified as field AGNs in the \textit{ROSAT} All-Sky Survey (RASS). 

By conducting an extensive follow-up survey of an all-sky X-ray point source catalog to look for galaxy overdensities, we will obtain a sample of such galaxy clusters. The primary question the sample will answer is whether there are other extreme-BCG clusters, similar to the Phoenix cluster, in our universe. This will tell us about the nature of highly efficient star formation in a galaxy cluster by distinguishing a short-lived phenomenon from a common occurrence in cool cores~\citep{2014McDonaldb}. Furthermore, we will identify any clusters with central QSOs as a secondary product of the survey. These clusters are also interesting in their own right, as they will allow us to study the effect of quasar-mode feedback on the ICM~\citep{2013Hlavacek}. This paper reports our first findings on this topic. Once the survey is complete, we will also have a better understanding of biases in X-ray surveys (i.e., the number of clusters missed due to the presence of a central point source), which is crucial for constraining cosmological parameters via cluster counts, such as the mean matter density $\Omega_m$, the normalization of the density fluctuation power spectrum $\sigma_8$, and the dark energy equation-of-state parameter $w_0$,~\citep{2008Mantz,2014Mantz,2009Vikhlininb}.

In order to reduce the total number of candidates to a manageable size, we require sources to be bright at X-ray, mid-IR, and radio wavelengths, relative to the optical. This requirement leads to a sample dominated by radio-loud type II QSOs and starbursts with central radio sources embedded in clusters. Subsequently, optical follow-up is performed to confirm the existence of a galaxy cluster via an overdensity of red galaxies at the same redshift as the central bright X-ray source. Results from the survey will be available in a forthcoming paper. Here, we focus on the first stage of the survey, including cross-correlation of all-sky surveys, and the first new discovery.

\subsection{Data Used in the Cross-correlation}\label{sec::data}
We expect that clusters with central QSOs or starbursts (or both) are extremely luminous at multiple wavelengths, including X-ray, mid-IR, and radio. Compact X-ray emission may be produced by the cooling ICM of a relaxed cluster or the hot accretion disk around a central AGN. Bright mid-IR emission traces warm dust which could be heated by a starburst and/or an AGN, while bright radio emission originates primarily in AGN jets, which are found ubiquitously in cool-core clusters~\citep{2009Sun}. We use the $K$ band as the \emph{normalization} and select relatively bright X-ray, radio, and mid-IR sources from that. The normalization is for preventing nearby sources (i.e., stars) from dominating the sample. Below, we describe how data at each of these wavelength is acquired.

\subsubsection{X-Ray Data: RASS} 
Our X-ray sample consists of 124,730 objects from the combination of the RASS Bright Source Catalog and Faint Source Catalog. RASS is the first all-sky survey in soft X-rays (0.1-2.4 keV), conducted in 1990/91 with ROSAT, a German X-ray telescope satellite~\citep{1999Voges}. However, this initial sample is dominated by sources that are not in clusters (i.e., isolated AGN, stars, etc) and requires additional cuts to reduce the total size to a manageable number for optical follow-up.

Several surveys in the past have used RASS to create X-ray flux-limited cluster catalogs. For example, the REFLEX survey has a flux limit of $3\times10^{-12}\,\rm{erg\,s^{-1}\,cm^{-2}}$~\citep{2004Bohringer} while the limit of ~\citet{2000Ebeling}'s Extended Brightest Cluster Sample was $2.8\times10^{-12}\,\rm{erg\,s^{-1}\,cm^{-2}}$ and that of~\citet{2001Ebeling}'s Massive Cluster Survey (MACS) faint extension was 1--2$\times10^{-12}\,\rm{erg\,s^{-1}\,cm^{-2}}$. We expect our survey to have a flux limit similar to the limites of these previous RASS-selected cluster surveys, given that we use the same data.

\subsubsection{Radio Data: NRVO Very Large Array Sky Survey (NVSS) and Sydney University Molonglo Sky Survey (SUMSS)}
Since there is no single radio all-sky survey available, the combination of two surveys, one for the northern hemisphere and one for the southern hemisphere, is necessary to achieve full-sky coverage for radio sources. For the northern hemisphere, the NVSS is a 1.4 GHz survey covering the entire sky north of a declination of $-40^{\circ}$~\citep{1998Condon}. For the southern hemisphere, the SUMSS was an 843 MHz survey covering the sky south of a declination of $-30^{\circ}$~\citep{2003Mauch}. Within the $2\sigma$ positional uncertainties from the X-ray and radio catalogs, we found 13,800 X-ray sources with a 1.4 GHz radio detection in NVSS or an 843 MHz detection in SUMSS. The given positional uncertainties account for the brightness of each source, and the systematic point-spread function (PSF) of the instruments. Because the two radio surveys do not cover the same wavelength, we scaled the flux from SUMSS to NVSS assuming a power-law spectrum from synchrotron radiation ($f_{\rm{NVSS}}/f_{\rm{SUMSS}} = (\rm{1.4}/\rm{0.834})^{-\alpha}$ where $\alpha$ = 1 is typical of radio galaxies in clusters~\citep{2015Hogan}). 

\subsubsection{Mid-infrared Data: Wide-field Infrared Survey Explorer (WISE)} 
\textit{WISE} was an all-sky survey with imaging capabilities at 3.4, 4.6, 12, and 22 $\mu m$~\citep{2010Wright}. We matched our 13,800 X-ray and radio sources to the AllWISE Source Catalog. Despite the fact that most of our sources have counterparts in \textit{WISE}, only $\sim50\%$ (7380 objects) of our sample has a $W4$ (22 $\mu m$) detection with a measurable $W4$ uncertainty, implying that only half of our X-ray and radio sources are relatively bright in mid-IR. The $W4$ band is used in this analysis because of its sensitivity to warm dust, heated by either star formations or AGN~\citep{2013Lee}.

\subsubsection{Near-infrared Data: Two Micron All Sky Survey (2MASS)} 
The 2MASS was a near-IR all-sky survey carried out with two automated 1.3 m telescopes, one in Arizona and one in Chile~\citep{2006Skrutskie}. The images were taken simultaneously at the $J$ (1.25 $\mu m$), $H$ (1.65 $\mu m$), and $K$ (2.17 $\mu m$) bands. As with WISE, we matched our X-ray and radio samples to the 2MASS All-sky Point Source Catalog (PSC) to extract the $K$-band brightness of our matched objects. We use the $K$ band because it is most sensitive to the stellar mass~\citep{Bell1999}. After cross-correlating WISE's $W4$ and 2MASS's $K$ band, we end up with 4549 targets for further follow-up.

\subsection{Color Cuts} \label{sec::method}
By requiring candidates to be detected in all four surveys (\textit{ROSAT}, NVSS or SUMSS, \textit{WISE}, and 2MASS), we are guaranteed soruces that are bright at all four wavelengths, which is a specific characteristics of a few astrophysical sources, including radio-loud type II QSOs (e.g., Fanaroff-Riley type I/II radio galaxies~\citep{1974Fanaroff}) and cooling-flow sources (e.g., the Phoenix cluster, the Perseus cluster, and Abell 1835). Since the total number of sources that exist in all four surveys remains too large (4549 objects) to perform the necessary follow-up, further cuts are required to identify the best candidates for our sample.

\begin{figure*}[!ht]
\begin{center}
\includegraphics[width=1.92\columnwidth]{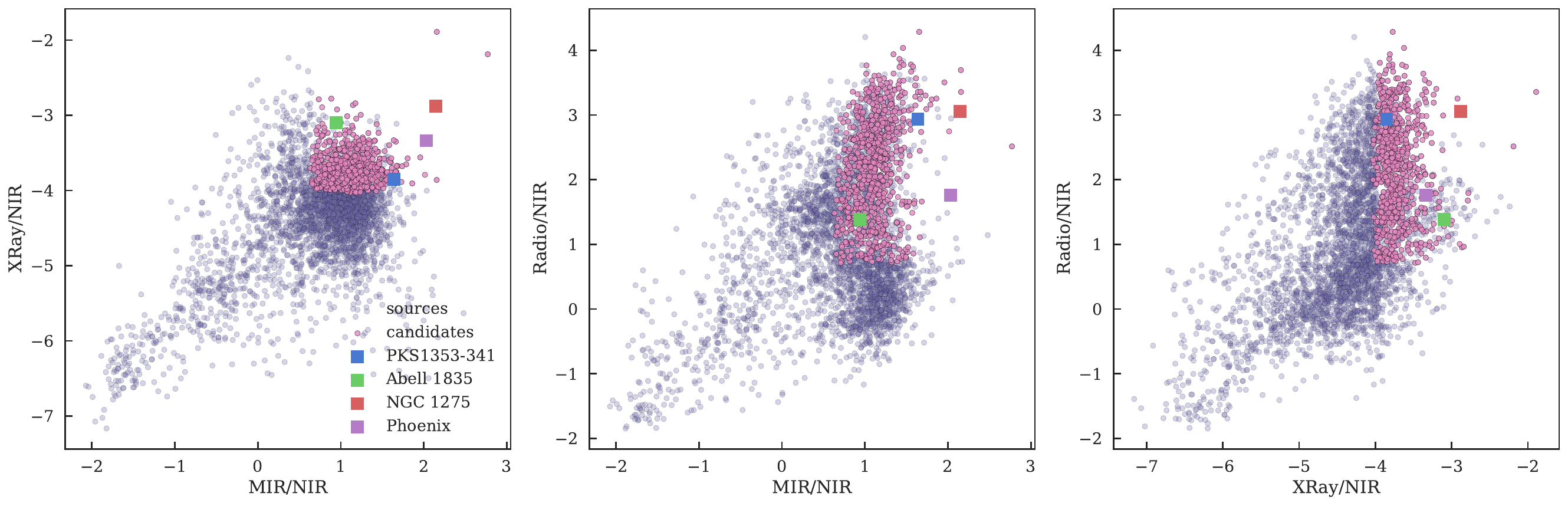}
\includegraphics[width=1.92\columnwidth]{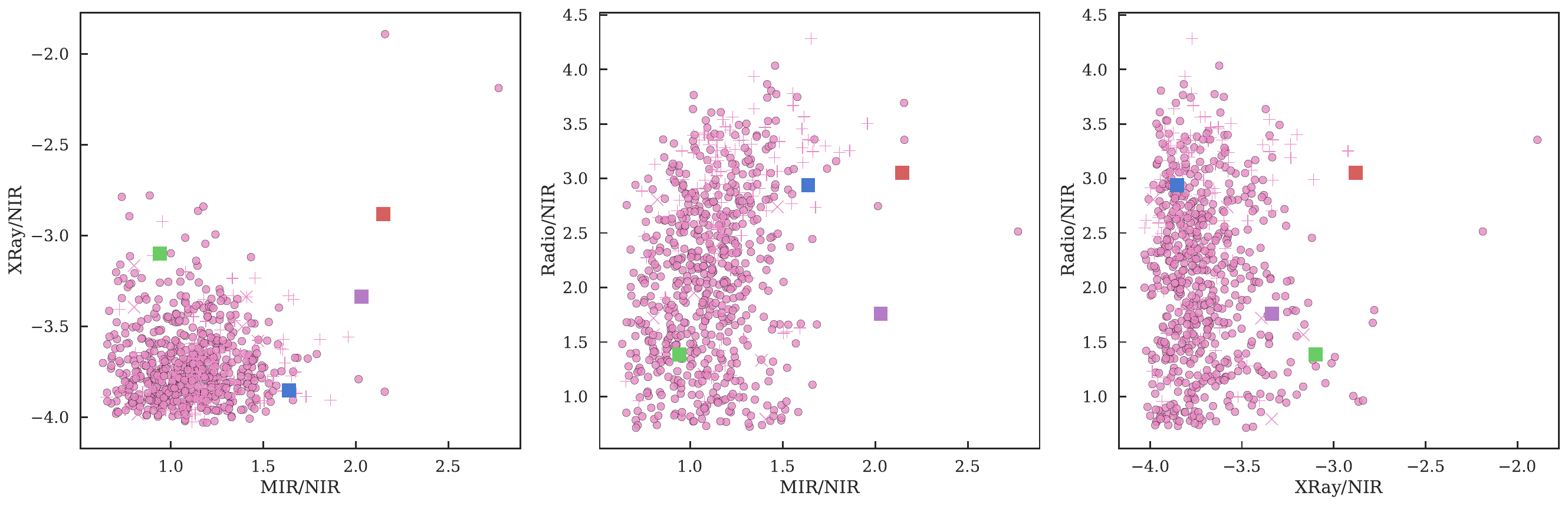}
\caption{Top three panels show color-color diagrams for objects that were detected in all four all-sky surveys (4,206 objects, see Section~\ref{sec::chips}). The axes are the logarithm of the ratio of the X-Ray, mid-IR (MIR) or radio flux to the near-IR (NIR) flux. Points in pink satisfy our three color cuts. The bottom three panels zoom in on these galaxy cluster candidates (735 objects). We remove the background ($z > 0.7$) and foreground ($z < 0.1$) sources from our sample based on redshift information from NED. The Phoenix, Perseus (NGC 1275), and A1835 clusters, which host extreme BCGs, are shown with purple, red and green squares, respectively while PKS1353-341 is shown with a blue square. }
\label{fig:color-cut}
\end{center}
\end{figure*}

After combining all four surveys, we start with a catalog of 4549 candidate clusters. The first cut is to remove stars from our local neighborhood by setting a $K$-band brightness threshold ($m_K>9$ mag). This reduces the number of candidates to 4,206. Subsequently, we apply a series of color-cuts at different wavelengths. In Fig.~\ref{fig:color-cut}, we plot the ratios of X-ray, mid-IR, and radio flux to near-IR flux. Normalizing to the near-IR flux takes into account each sources's overall brightness, which strongly depends on the source's distance. We selected sources from the top right corner of each plot, i.e., objects that are over-luminous in X-ray, radio, and mid-IR compared to near-IR. As a result, we reduce our sample from 4206 to 735 objects. Specifically, the regions of the color cut for X-ray, radio, and mid-IR were chosen to have their minimum flux normalized to near-IR, lower than that of the Phoenix cluster by $\log_{10}3$, $\log_{10}9$, and $\log_{10}15$, respectively. These ratios were obtained by considering the expected range of color for a Phoenix-like object at an unknown redshift between 0.1 and 0.7. The 735 remaining objects define our primary sample.

For 428 out of our 735 cluster candidates, we obtain the redshift for the bright source from the \textit{NASA/IPAC Extragalactic Database} (NED)\footnote{https://ned.ipac.caltech.edu}. We rejected foreground (redshift less than 0.1) and background (redshift greater than 0.7) objects. For clusters at $z < 0.1$, diffuse emission should be readily detected by eye even in the presence of a bright central point source. Thus, we do not expect many clusters to have been missed at these redshifts. At $z > 0.7$, cluster detection in the optical becomes challenging because of the limitations in detecting the red sequence from ground-based telescopes~\citep{2000Gladders}, and we are likely to miss them in our shallow all-sky survey data. This challenge, along with follow-up efforts, will be addressed in a forthcoming paper.

\begin{figure*}[!ht]
\begin{center}
\includegraphics[width=2.\columnwidth]{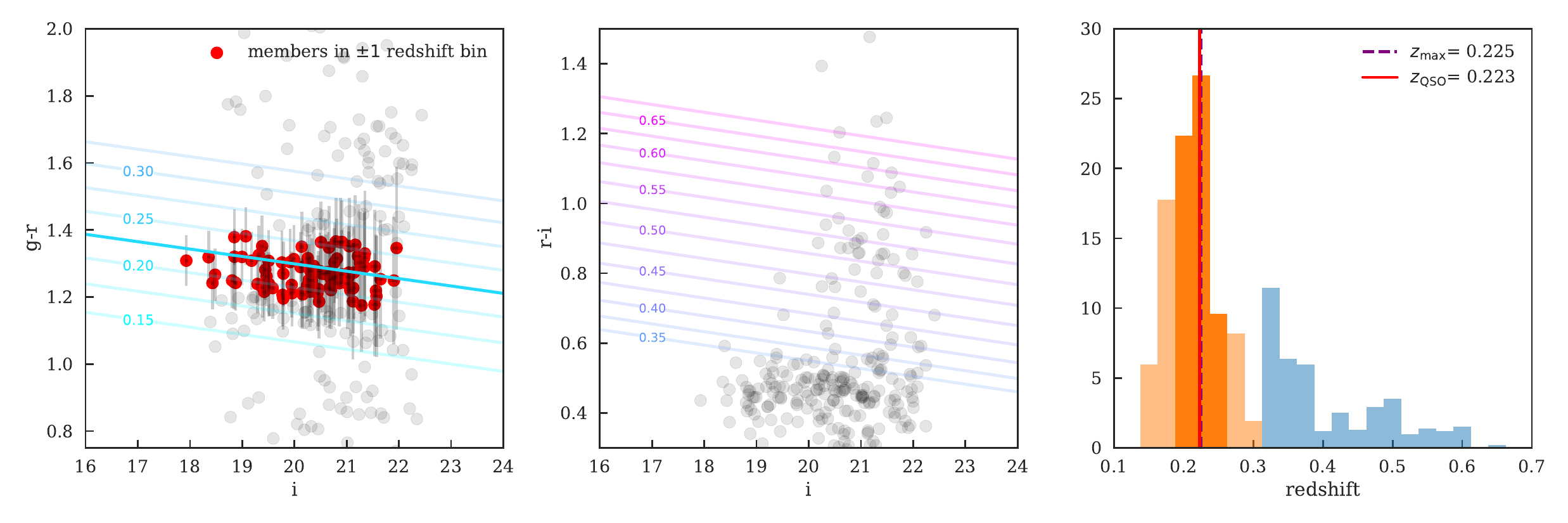}
\caption{Left and Center: Color magnitude diagrams for PKS1353-341 using $g-r$ color (left) and $r-i$ color (center) to identify the red sequence at a variety of redshifts (diagonal lines).
All the members within $\pm1$ redshift bin of the QSO's redshift are shown in red dots. Right: The number of galaxies in different redshift bins based on the red sequence templates shown in the left panels. The g-r vs i color-magnitude diagram (orange bins) is used for $z=0.15\textendash0.35$ and the r-i vs i color-magnitude diagram (blue bins) is used for $z=0.35\textendash0.70$. The redshift of the quasar, PKS1353-341, is 0.223 in red solid line while the redshift of the maximum histogram is 0.225 in purple dash line. This figure demonstrates how we can use the red sequence to find new clusters from X-ray point source catalogs up to redshift $\rm{z} = 0.7$.}
\label{fig:red_sequence}
\end{center}
\end{figure*}
As a pilot study of the CHiPS survey, we selected 22 candidates, which were both Phoenix-like (top right corner in Fig.~\ref{fig:color-cut}) and visible to observe from the 6.5-meter Magellan telescope in the spring of 2014. These candidates were initially imaged with the Inamori Magellan Areal Camera and Spectrograph~\citep{2011Dressler} on the Magellan Baade telescope, and then promising candidates were further imaged using the Parallel Imager for Southern Cosmological Observations \citep[PISCO;][]{2014Stalder}, to a depth sufficient to detect red-sequence galaxies at $z\sim0.6$. PISCO is a photometer that produces $g$, $r$, $i$, and $z$ band images simultaneously within a 9$^{\prime}$ field of view. Creating four band images at the same time increased our efficiency in observing these candidates by a factor of $\sim$3 (including optical losses). Further discussion about the reduction pipeline will be made in an upcoming paper. With the optical images obtained from PISCO, we searched for an overdensity of red sequence galaxies~\citep{2000Gladders}, selecting candidates which have a significant ($>3\sigma$) overdensity of red galaxies at the same redshift as the cluster candidate. This led to an initial sample of four galaxy cluster candidates, which were followed up with the \textit{Chandra} X-ray telescope. This follow-up resulted in the discovery of a new massive galaxy cluster surrounding PKS1353-341 at $z = 0.223$ with R.A. = $\rm{13^h56^m05.4^s}$ and decl. = $\rm{-34^d21^m10.9^s}$. The red sequence of PKS1353-341, shown in the right panel of Fig.~\ref{fig:red_sequence}, demonstrates that the redshift of the QSO is similar redshift to that of the maximum histogram bin for the red sequence. This suggests that most of the surrounding red galaxies are located near the QSO in the physical space, and not in projection. Fig.~\ref{fig:red_sequence} also demonstrates the capability of this technique to detect galaxy clusters using just optical photometry from three bands ($g$, $r$, $i$) up to redshift $z=0.7$.

Two of the other three candidates turn out to be isolated X-ray point sources, implying that the galaxy overdensity exists only in projection. This led us to refine our selection algorithm which will be presented in detail in Somboonpanyakul et al. (in prep). The remaining candidate is a rich cluster (A2270) with no existing \emph{Chandra} data. In our observations it clearly shows extended X-ray emission.

\section{Reduction of \textit{Chandra} Data} \label{sec::chandra}
To confirm the existence of a massive galaxy cluster, X-ray observation is important as it provides unambiguous evidence for an extremely hot ICM, which is expected from the deep potential well of a cluster. In particular, high angular resolution X-ray images can be used to determine different properties of this hot ICM, such as the gas temperature profile, gas density profile, and total hydrostatic mass. 

\begin{figure}[!ht]
\begin{center}
\includegraphics[width=0.48\columnwidth]{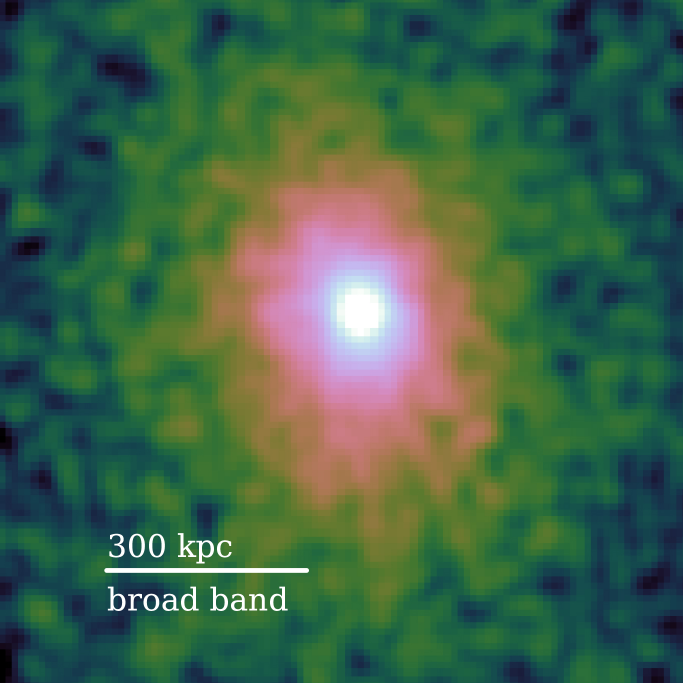}
\includegraphics[width=0.48\columnwidth]{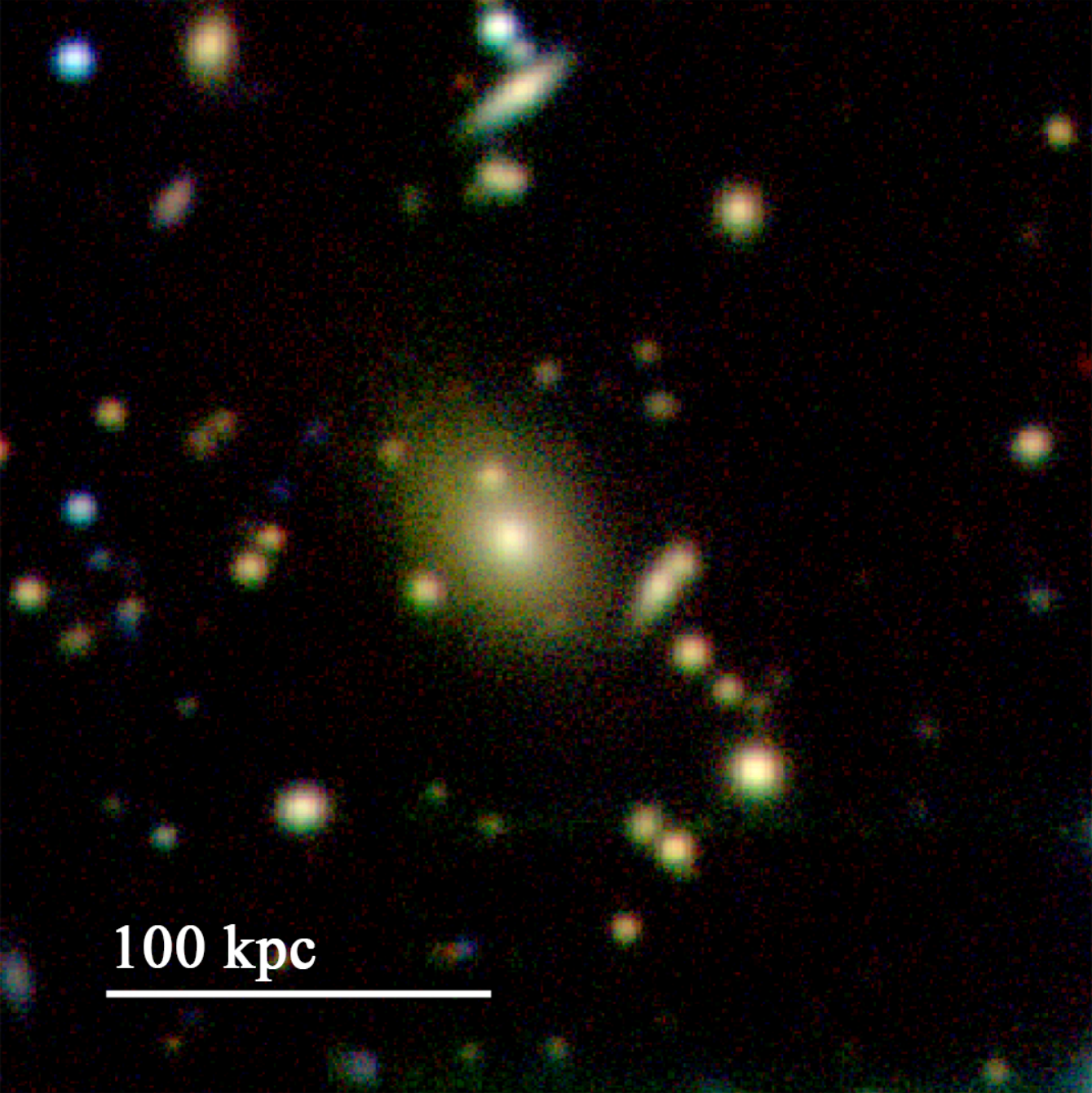}
\caption{Left: \textit{Chandra} broadband (0.5-7 keV) image for PKS1353-341 on a log-scale color bar, showing the bright central point source and the surrounding diffuse cluster emission. Right: Magellan {PISCO} ($g$ , $r$ , $i$) image of the inner part of the galaxy cluster, showing the central giant elliptical galaxy.}
\label{fig::image_full}
\end{center}
\end{figure}

Fig.~\ref{fig::image_full} shows both X-ray and optical images of PKS1353-341. The optical image from the Magellan telescope (the right panel of Fig.~\ref{fig::image_full}) shows the central BCG as an extremely bright elliptical galaxy with a bright point source in the center and a number of elliptical members nearby. The smoothed \textit{Chandra} X-ray image clearly shows the extended hot ICM, which reveals the morphology of the cluster to be highly relaxed without obvious perturbation in any direction, implying that this cluster has not experienced any recent mergers. The image also reveals a central concentration, apart from the central point source, reminiscent of a cool core. 

In the following sections, we describe the reduction of the \textit{Chandra} data, followed by derivation of various ICM properties.

\subsection{Data Preparation}
PKS1353-341 (OBSID 17214) was observed with \textit{Chandra} ACIS-I for 31 ks. The cluster has a bright point source at the center that is not piled up. Excluding the point source, the total number of counts we used for the reduction is 22,258. It was analyzed with \textit{CIAO} version 4.8 and \textit{CALDB} version 4.7.2, provided by the \textit{Chandra} X-ray Center. The event data was recalibrated with VFAINT mode to improve background screening. Point sources not at the center of the cluster were identified using the \textit{wavdetect} function and removed. A blank background was generated using the \textit{blanksky} script, which includes combining and reprojecting the background for the input event file. Spectra were extracted in concentric annuli (defined below), for both the cluster and background, using the \textit{specextract} function from \textit{CIAO}.

In order to get reliable measurements of the gas properties, such as density and cooling time, near the center of the cluster, a clear separation between the central point source and the galaxy cluster is required. The \textit{Chandra} PSF is complicated due to smearing effects from the High-Resolution Mirror Assembly, which are produced by a combination of telescope dithering motion, the size of detector pixels, and detector effects\footnote{http://cxc.harvard.edu/ciao/PSFs/psf\_central.html}. We generate a simulation of the central point source, described below, to account for and remove the contribution of the point source to the total emission, following~\citet{2010Russell}.

\subsection{Simulating quasar PSF with ChaRT} \label{sec::chart}
ChaRT~\citep{2003Carter} is a web interface\footnote{http://cxc.harvard.edu/ciao/PSFs/chart2/runchart.html} to the SAOTrace ray trace code for creating a simulated point source from a given source spectrum. In order to properly use ChaRT, we must prepare ChaRT inputs, including the source spectrum, source coordinates, and pointing information for the telescope. The pointing information is acquired from the aspect solution of the \textit{Chandra} image, while the source spectrum was obtained from \textit{specextract} for a region dominated by the point source (central). 

We use Sherpa~\citep{2001Freeman} to model the X-ray emission from the inner $2^{\prime\prime}$, assuming a combination of an absorbed power-law model to represent the central AGN ($xszphabs \times xspowerlaw$) and a thermal plasma model to represent the ICM ($xspec$), with photoelectric absorption from the Milky Way ($xsphabs$) applied to both components. The hydrogen absorbing column ($n\rm{_H}$) was fixed to the average from the Leiden-Argentine-Bonn survey at $5.57\times10^{20}$ cm$^{-2}$~\citep{2005Kalberla}. 

The output of ChaRT is a set of simulated rays from the point source. The MARX software (version 5.3) projects the simulated rays onto the detector plane to create pseudo-event files. Instead of using MARX directly, we utilized \textit{CIAO}'s $simulate\_psf$ script to simplify its interface. The surface brightness profile of the simulated point source was extracted according to the procedure described in Section~\ref{sec::density}. 

\subsection{Density Profile} \label{sec::density}
Radial gas density profiles were created by first obtaining the number of counts at 0.5-2 keV in concentric annuli and dividing this number by the annulus area to get the surface brightness in count per square arcsecond. The spacing for the radial bins is equally separated in the logarithmic scale for 30 bins with a minimum spacing of 1$^{\prime\prime}$ and excluding the central 1$^{\prime\prime}$ where the central point source dominates. We use a maximum radius of $\sim$450$^{\prime\prime}$ to ensure that we have enough counts to get a good constraint on the surface brightness for each annulus. The simulated point source profile was subtracted from the surface brightness profile, as described in the previous section, to remove the surface brightness contribution from the central AGN, (see Fig.~\ref{fig::17214_temp}). The surface brightness was converted to units of physical density using the normalization terms from the spectral fitting (see Section~\ref{sec::temp}).

The analytic expression for the 3D density profile that we use, a modified $\beta$-model, represents features of observed X-ray density profiles, including a power-law cusp, a two-component $\beta$-function with small- and large-scale slopes, and a second $\beta$-model component with a small core radius to represent the cool core~\citep{2006VikhlininA}. The complete model for the density profile is
\begin{align} \label{eqn:density}
n_p n_e(r) = & n_0^2 \frac{(r/r_{c})^{-\alpha}}{(1+(r/r_{c})^2)^{3\beta-\alpha/2}}\frac{1}{(1+(r/r_{c})^\gamma)^{\epsilon/\gamma}}\\
& +\frac{n_{02}^2}{(1+(r/r_{c2})^2)^{3\beta}} \nonumber,
\end{align}
where $n_p n_e(r)$ is the product of the proton and electron densities. We fixed $\gamma=3$ and $\epsilon\le5$, while all other parameters were free. Before fitting the 3D model to the 2D data, we project the model onto a 2D plane by integrating along the line of sight~\citep{2006VikhlininA}. This projected model is fit to the model, using \textit{emcee}~\citep{2013Foreman-Mackey}.

\subsection{Temperature Profile} 
\label{sec::temp}
We extract spectra in coarser annuli so that the number of counts per annulus is roughly 1500, to allow for well-constrained temperature measurements. The first annulus has an inner radius of 1.$^{\prime\prime}$5 (5 kpc) to avoid contamination from the AGN. We extract the cluster and blank-sky background spectra for each annulus. All spectra were fit simultaneously with APEC models for both the cluster (\textit{xsapec}) and the Milky Way emission (\textit{xsapec$_{\rm{bkg}}$}), photoelectric absorption from the Milky Way (\textit{xsphabs}), and thermal bremsstrahlung from unresolved background objects (\textit{bremss}) with a temperature of 40 keV, following~\citet{2014McDonald}. The temperature, metallicity, and normalization of the cluster emission model were left free and the WSTAT statistic was used. This produces a temperature profile with 7 bins over roughly the inner 700 kpc.

Instead of fitting a model with many free parameters to the poorly-constrained temperature profile, we leverage the high-S/N density profile with the fact that the pressure profile of a galaxy cluster is close to self-similar and shows little scatter at large radii~\citep{2007Nagai,Arnaud2010, 2014McDonald}. The temperature profile is then inferred from the ideal gas law ($P=n_ekT$) and the density profile. We model the pressure profile with a modified generalized Navarro-Frenk-White (GNFW) model, as proposed by~\citet{2007Nagai}:
\begin{equation}\label{eqn::press}
P(r) = \frac{P_0}{x^\gamma[1+x^\alpha]^{(\beta-\gamma)/\alpha}},
\end{equation}
where $x=r/r_s$, and $\alpha$, $\beta$, and $\gamma$ are the slopes at $r\sim r_s$, $r\gg r_s$, $r\ll r_s$, respectively. Slope $\beta$ was fixed at 5.4905~\citep{Arnaud2010}, leaving 4 free parameters ($P_0$, $\alpha$, $\gamma$, $r_s$). 

Dividing the model pressure profile by the model density profile yields a 3D model temperature profile which was then projected along the line of sight, using the weighting scheme for the average temperature proposed by~\citet{2004Mazzotta}:
\begin{equation}\label{eqn:T2d}
\langle T(r) \rangle = \frac{\int_V \rho^2 (T(r))^{1/4}d^3x}{\int_V \rho^2 (T(r))^{-3/4}d^3x},
\end{equation}
where $\rho$ is the gas density, $T(r)$ is the 3D temperature profile and $\langle T(r) \rangle$ is the projected 2D temperature profile at a given radius. The projected temperature profile, which has three free parameters from the GNFW model, is then fit to the data, using \textit{emcee}~\citep{2013Foreman-Mackey}.

\subsection{Total Mass, Gas Fraction, Entropy, and Cooling Time} \label{sec::others_method}
Having both the density and temperature profiles allows us to calculate other thermodynamic properties of the cluster, including the enclosed total mass as a function of distance, assuming hydrostatic equilibrium~\citep{Sarazin2009},
\begin{equation} \label{eqn::mass}
M(r)=-\frac{k T_g (r) r}{\mu m_p G}\left(\frac{d\ln{P}}{d\ln{r}}\right)
\end{equation}
where $P$ is the gas pressure, $T_g$ is the cluster emission temperature, $\mu$ is the chemical abundance (which is equal to 0.5954 for primordial He abundance), and $m_p$ is the proton mass. We choose $M_{500}$--the total mass within $R_{500}$, the radius within which the average enclosed density is 500 times the critical density $\rho_c \equiv 3H_0^2/8\pi G$--as a proxy for the total cluster mass, as suggested by~\citet{2009Vikhlinin}.

The gas mass of the cluster can be calculated by integrating the gas density over a spherical volume using $M_{gas}(r)=\int^r_0 \sqrt{n_pn_e(r)} \times 1.276 \times m_pdV$, where $n_p n_e(r)$ is the product of the proton and electron densities, and $1.276$ is the ratio of protons to electrons in a plasma with 0.3 $\rm{Z}_{\odot}$ metallicity. From this, the gas fraction interior to some radius is calculated by $f_{gas}(r)=\frac{M_{gas}(r)}{M(r)}$.

With the density and temperature profile, the entropy of the ICM can be calculated using $K(r)=kT(r)\times n_e(r)^{-2/3}$, where $kT(r)$ is the temperature profile (in kiloelectronvolts) and $n_e(r)$ is the electron density profile. Entropy is a useful observable for studying the effects of feedback on a cluster because the thermal history of a cluster is influenced solely by heat gains and losses~\citep{2000Lloyd-Davies, 2009Cavagnolo, 2013Panagoulia}. We expect a monotonically increasing entropy profile due to the buoyancy of high-entropy gas~\citep{2005Voitb,2009Cavagnolo}.

Lastly, the cooling time represents the amount of time that the ICM needs to radiate all of the excess heat via thermal bremsstrahlung emission. This is calculated using $t_{cool}=\frac{kT(r)}{n_e(r)\Lambda(T)}$, where $T(r)$ is the temperature profile, $n_e(r)$ is the electron density profile and $\Lambda(T)$ is the cooling function~\citep{1993Sutherland}. The central cooling time (the cooling time within the central $\sim\!\!10$ kpc) is often used to distinguish between cool core and non-cool core clusters~\citep{2010Hudson}.

\section{Results} \label{sec::result}
The fact that this galaxy cluster was not identified by \textit{ROSAT} as a cluster suggests that there may be a hidden population of galaxy clusters hosting extreme central galaxies (i.e., starbursts and/or QSOs). The unabsorbed bolometric X-ray flux of the cluster is $4.8\times10^{-12}\,\rm{erg/cm^2/s}$ and the bolometric X-ray luminosity at $z=0.223$ is $L_x\sim7\times10^{44}$ $\rm{erg\,s^{-1}}$. The bolometric X-ray lumnisity of the core is $L_{nuc}\sim1.8\times10^{44}$ $\rm{erg\,s^{-1}}$.
 
Table~\ref{table::keyvalue} shows the key properties of PKS1353-341, which are derived in this work ($R_{500}$, $M_{500}$, $M_{\rm{gas},500}$, $T_x$, $t_{\rm{cool},0}$, SFR), compared to other well-known clusters--namely, A1795 (a strong cool core (SCC) cluster) and H1821+643 (a quasar-hosting cluster). This table shows that PKS 1353-341 is very similar to A1795 (a typical relaxed cool-core cluster), except in terms of its central AGN.

\begin{deluxetable*}{cccc}
\tabletypesize{\footnotesize} 
\tablecaption{Key properties for the galaxy cluster\label{table::keyvalue}}
\tablecolumns{0}
\tablewidth{0pt} 
\tablehead{ \colhead{Property of Clusters} & \colhead{PKS1353-341} & \colhead{A1795\tablenotemark{d}} & \colhead{H1821+643\tablenotemark{e}}} 
\startdata
Redshift\tablenotemark{a} ($z$) & 0.2230 & 0.0622 & 0.299 \\
$T_x$\tablenotemark{b} (keV) & $4.32\substack{+1.74\\-1.92}$ & $6.12\pm0.05$ & 8.9 $(0.15-0.75R_{500})$ \\
$R_{500}$ (kpc) & $1313\substack{+230\\-194}$ & $1235\pm36$ & $1000$ \\
$M_{500}$ ($10^{14}\,M_{\odot}$) & $6.90\substack{+4.29\\-2.62}$ & $6.03\pm0.52$ & $9$ \\
$M_{\rm{gas},500}$ ($10^{13}\,M_{\odot}$) & $6.45\substack{+1.41\\-1.22}$ & $6.27\pm0.65$ & $13$ \\
$f_{\rm{gas},500}$ & $0.094\substack{+0.042\\-0.018}$ & $0.104\pm0.006$ & 0.14\\
$t_{\rm{cool,0}}$ (Myr) & $299\substack{+92\\-70}$ (at 10 kpc) & 889 (at 10 kpc) & $1000$ (at 30 kpc) \\
$r_{\rm{cool}}$\tablenotemark{f} (kpc) & $185\substack{+12\\-11}$ & 82 & 90 \\
SFR\tablenotemark{c} ($M_\odot\,\rm{yr^{-1}}$) & $6.2\pm3.6$ & $9$ & $300\substack{+300\\-200}$ \\ Cooling rate\tablenotemark{f} ($M_\odot\,\rm{yr^{-1}}$) & $345\substack{+41\\-37}$ & 294 & $300\pm100$ \\
\enddata\tablenotetext{a}{Redshift is obtained from \textit{NED}. We assume that the cluster is located at the same redshift as the central AGN.}
\tablenotetext{b}{$T_x$ is measured from 0.15$R_{500}$ to 1.0$R_{500}$.}
\tablenotetext{c}{SFR is measured from the UV luminosity of the BCG for PKS1353-341. (see Section~\ref{sec::sfr})}
\tablenotetext{d}{Most of the numbers for A1795 are from~\citet{2006VikhlininA}, except SFR is from~\citet{2005Hicks}. $t_{\rm{cool,0}}$, $r_{\rm{cool}}$, and cooling rate are from~\citet{McDonald2017}}

\tablenotetext{e}{These numbers are from~\citet{2010Russell,2014Walker}, except SFR which is from~\citet{2013Ruiz}.}
\tablenotetext{f}{The cooling radius is defined to be the radius at which the cooling time is 7.7 Gyr while the cooling rate is defined within the cooling radius.}
\end{deluxetable*}

In the following sections, we discuss in more details various derived properties of the cluster, including the gas fraction, the entropy, the total hydrostatic mass, and the cooling time. 

\subsection{Temperature and Density profile} \label{sec::tempprofile}

\begin{figure*}[!ht]
\begin{center}
\includegraphics[width=1.0\columnwidth]{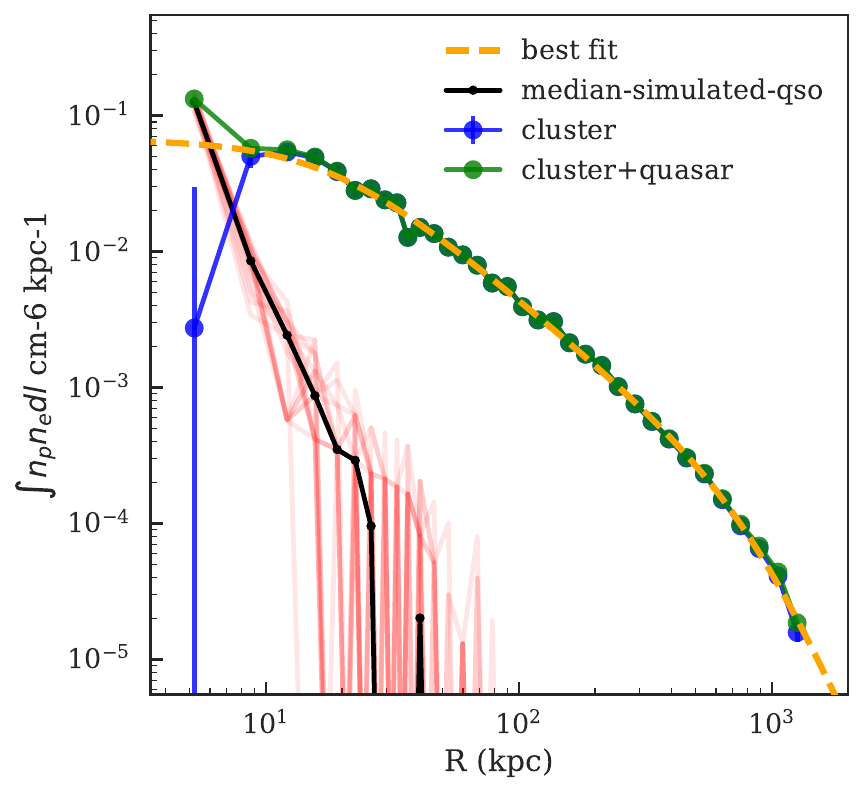}
\includegraphics[width=0.95\columnwidth]{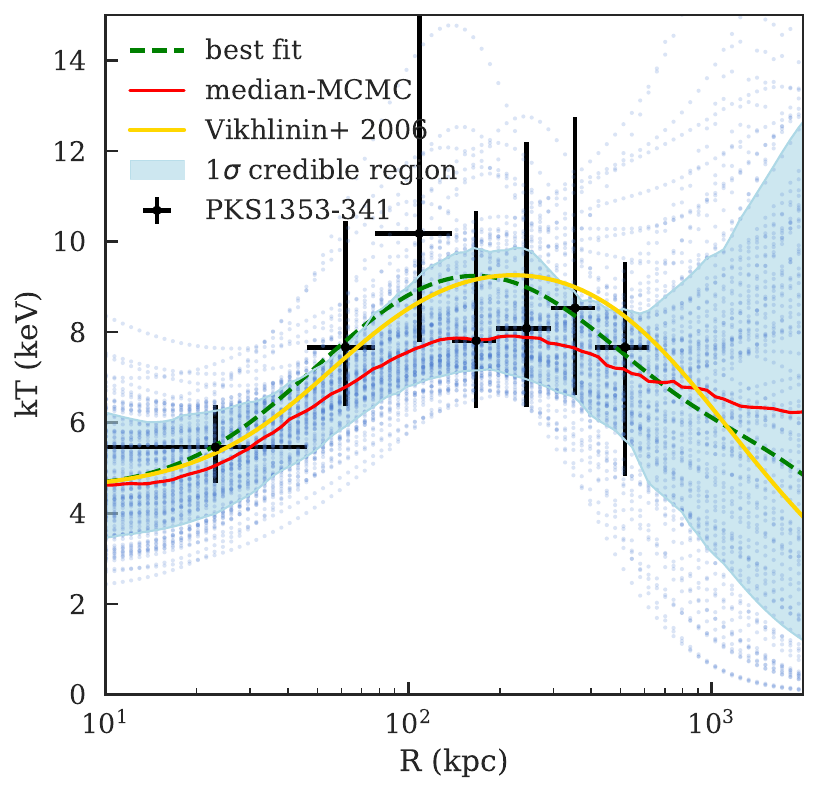}
\caption{Left: the surface brightness profile with the total brightness in green and with the removed simulated point source in blue. The red lines are different realizations of the simulated point source, while the black line is their median. The orange line is the best-fit model. Right: the temperature profile of the cluster. The blue dotted lines are different Markov-chain Monte Carlo (MCMC) realizations for the fit, using \textit{emcee}, a Python ensemble sampler~\citep{2013Foreman-Mackey}. The cyan-shaded region corresponds to a $1\sigma$ credible region from the MCMC simulation. The green dashed line is the best-fit model, and the red solid line is the median of the MCMC. The yellow line is the universal temperature profile from~\citet{2006VikhlininA}.}
\label{fig::17214_temp}
\end{center}
\end{figure*}

As shown in the left panel of Fig.~\ref{fig::17214_temp}, the cluster has a density profile with a sharp peak in the innermost bin (the central 1$^{\prime\prime}$). From the simulated PSF in Section~\ref{sec::chart}, an AGN surface brightness profile is created, shown by the black solid line. This simulated profile was subtracted from the observed profile, leaving the underlying ICM density profile. The green points represent the cluster density profile after the subtraction of the simulated AGN profile. Due to PSF modeling uncertainties, the innermost bin has a large residual uncertainty. However, the overall ICM density profile fit is relatively insensitive to the first data point. 

Based on the temperature profile in the right panel of Fig.~\ref{fig::17214_temp}, the cluster appears to harbor a cool core. Its temperature profile rises sharply from a minimum of 5 keV at $\sim$10 kpc to a maximum of 10 keV at $\sim$100 kpc with a core-excised ($0.15\textendash1.0\,\rm{R_{500}}$) temperature of $4.32\substack{+1.74\\-1.92}\,\rm{keV}$. The temperature profile is not well-constrained at radii greater than 700 kpc. In addition, the temperature profile has a similar shape to the universal profile (thick yellow line) derived by~\citet{2006VikhlininA}.

\subsection{Total Mass, Gas Mass Fraction, Entropy, and Cooling Time} \label{sec::others}
The total hydrostatic mass within R$_{500}$ is $M_{500}=6.90\substack{+4.29\\-2.62}\times10^{14}\,M\rm{_\odot}$. For comparison, the total mass within $R_{500}$ of A1795 is $(6.03\pm0.52)\times10^{14}\,M\rm{_\odot}$, while that of H1821+643 is $9\times10^{14}\,M\rm{_{\odot}}$~\citep{2006VikhlininA,2010Russell}. We measure a gas fraction of $\sim$0.1, consistent with what is found within $R_{500}$ for typical clusters~\citep{Allen2008,2013Gonzalez}.

\begin{figure*}[!ht]
\begin{center}
\includegraphics[width=1.95\columnwidth]{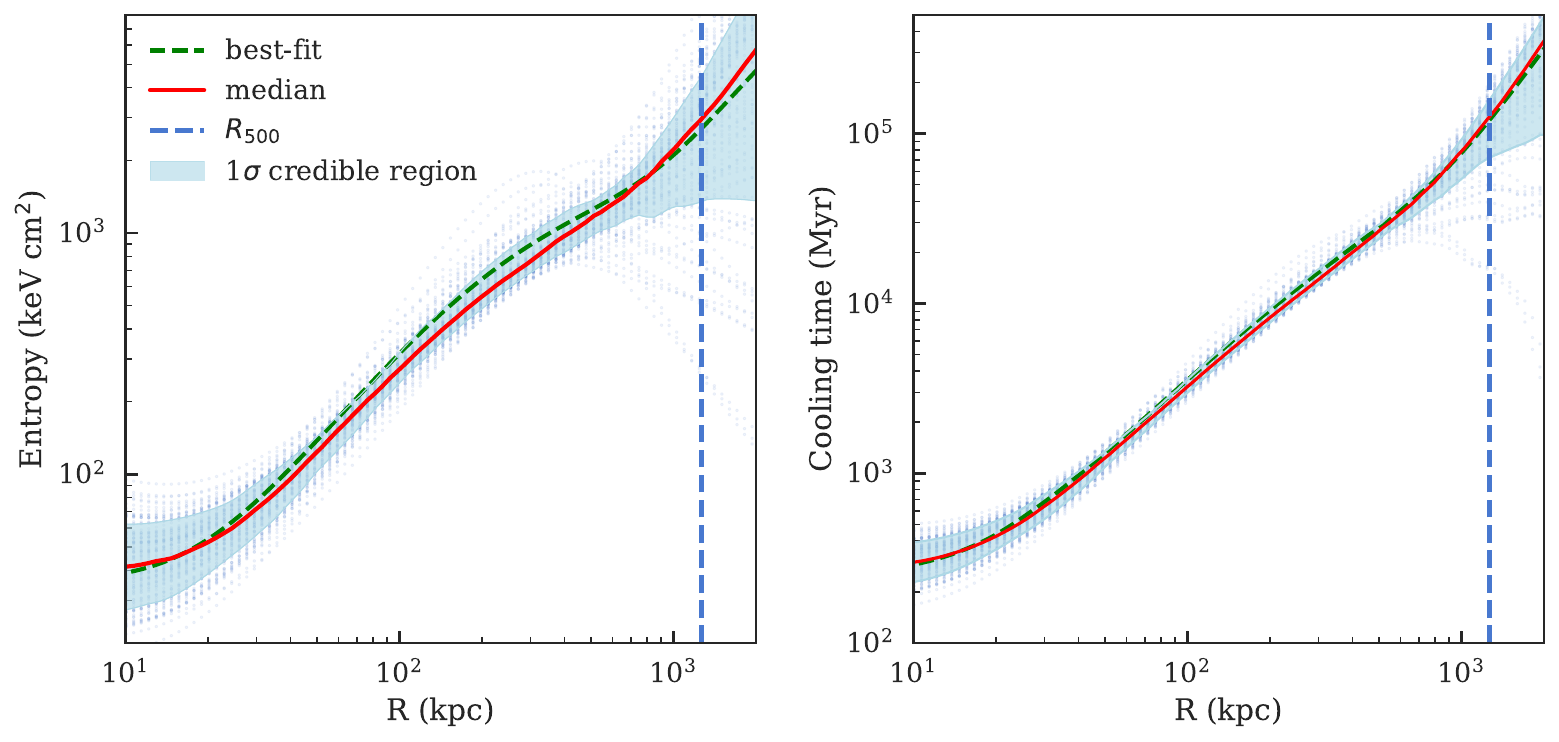}
\caption{Left: entropy profile for PKS1353-341. Right: cooling time profile for PKS1353-341. The best-fit-parameter model is displayed as a green dashed line and the median model from MCMC simulation is displayed as a red solid line. The cyan-shaded region corresponds to a $1\sigma$ credible region from the MCMC simulation, the blue dotted lines are different realization of the MCMC, and the blue dashed line is an $R_{500}$ for the cluster (1264 kpc).}
\label{fig::17214_entropy}
\end{center}
\end{figure*}

The entropy profile for this cluster is shown in the left panel of Fig.~\ref{fig::17214_entropy}. At all radii, the entropy profile is monotonically increasing. The best fit $K_0$, the core entropy, is $\sim50\,\rm{keV\,cm^2}$, which leans toward the smaller-value mode for the bimodal distribution of $K_0$, found by~\citet{2009Cavagnolo}. A discussion of this entropy profile in the context of other galaxy clusters will be presented in Section~\ref{sec::dissen}.

Lastly, the right panel of Fig.~\ref{fig::17214_entropy} shows the cooling time profile of the cluster. The cooling time profile has a monotonically increasing profile from the center of the cluster to larger radii. The central cooling time (at 10 kpc from the center) of PKS1353-341 is $299\substack{+92\\-70}\,\rm{Myr}$. According to~\citet{2010Hudson}, if the central cooling time is less than 1 Gyr, then the cluster is classified as an SCC cluster. Therefore, this cluster is an SCC cluster. Both the temperature and entropy profiles also lead to the same conclusion. 

The radius of the cooling region, defined as the radius at which the cooling time is less than 7.7 Gyr, was also measured and found to be $\sim$200 kpc. Without any source of feedback, the ICM inside this radius should have cooled since $z\sim1$, which suggests that some heating mechanism is required to keep the ICM hot on scales of $\sim$200 kpc~\citep[i.e., AGN feedback;][]{2014McNamara}.

\section{Discussion} \label{sec::diss}
Close examination of PKS1353-341 reveals that it has similar physical properties to other, well-studied galaxy clusters, including its entropy profile and its total enclosed mass. In this section, we discuss the cluster in a larger context and explore other aspects of the cluster, such as the BCG star formation rate (SFR) and the AGN feedback. 

\subsection{Entropy Profile} \label{sec::dissen}
\begin{figure}[!ht]
\begin{center}
\includegraphics[width=0.95\columnwidth]{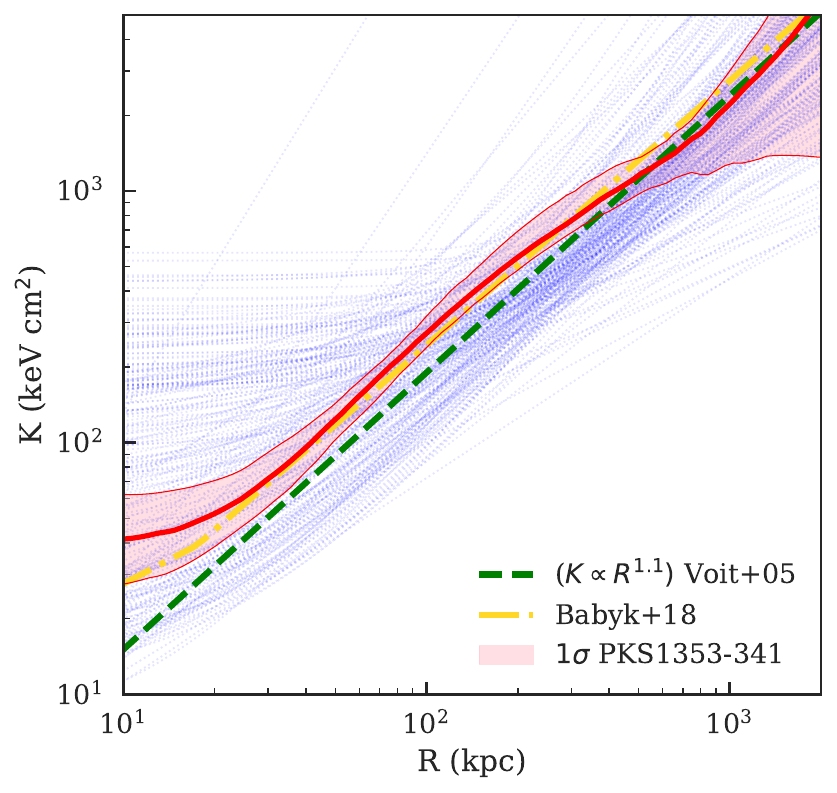}
\caption{Entropy profile for PKS1353-341, compared to 239 clusters from \citet{2009Cavagnolo} in blue. Best-fit entropy profiles from \citet{2005Voitb} and \citet{Babyk2018} are shown in green and yellow, respectively. Despite having one of the lower entropies at 10 kpc, PKS1353-341 has one of the highest entropies at 200 kpc. This bump may be due to a past energetic event at the cluster center depositing heat, which has been rising buoyantly.}
\label{fig::entropy_prof}
\end{center}
\end{figure}

The behavior of the entropy profile for this cluster, as described in Section~\ref{sec::others}, is consistent with what~\citet{2009Cavagnolo} found in his entropy profile analysis of the ICM for 239 clusters in the \textit{Chandra} archive. The blue dotted lines in Fig.~\ref{fig::entropy_prof} show all 239 clusters in Cavagnolo's sample, and the red solid line is that of PKS1353-341. The two dashed lines are for a simple power law model based on~\citet{2013Panagoulia} and~\citet{2017Hogan}.

The central ($r<10$ kpc) entropy of the cluster is approximately $30\textendash60\,\rm{keV\,cm}^2$. The large uncertainty comes from the uncertainty of the core density and temperature, due to the central AGN. The nonzero core entropy can be explained by either the AGN providing a large amount of energy to offset cooling and maintain nonzero entropy~\citep{2005Voit} or the low-entropy gas near the core having cooled and the core being refilled with higher-entropy gas. We see no significant flattening of the entropy profile near the center of the cluster, similarly to what was suggested by~\citet{2013Panagoulia},~\citet{2017Hogan} and~\citet{Babyk2018}.

One surprising result of the entropy profile is an excess above a power law model at $\sim$200 kpc. Quantitatively, the entropy of PKS1353-341 in the core is lower than the median ($\sim$35th percentile) of the 239 clusters from~\citet{2009Cavagnolo}, consistent with being a cool-core cluster. At large radii ($\sim$750 kpc), the measured entropy for PKS1353-341 is consistent with the median value for the 239 clusters, demonstrating the self-similar nature of clusters outside of the core. However at $\sim$200 kpc, the entropy for PKS1353-341 is 1$\sigma$ above the median ($\sim$84th percentile), implying that the excess is marginally significant. Given that it is only a 1$\sigma$ deviation, this could very well be simply a statistical fluctuation, or a result of model assumptions. However, if the excess is real, this behavior is unusual because any excess high-entropy gas should rise to larger radius, smoothing out any departure from a power law in relatively short timescale. Thus, any large-scale deviation from this profile is likely caused by some transient non-gravitational process in the core, such as AGN heating~\citep{2005Voitb,2013Panagoulia}. One possible explanation for the excess at $\sim$200 kpc is that the AGN recently heated up the low-entropy gas at the core, which rises to a greater radius (i.e., at $\sim$200 kpc). As the heated gas moves outward, the high-entropy gas starts to mix with colder surrounding gas, which smooths out the excess entropy. But in this case, there is not enough time for this process to complete, leaving the excess at $\sim$200 kpc. Assuming that excess entropy deposited at the center rises at the sound speed, the time it takes to reach a radius of 200 kpc in PKS1353-341 is the sound crossing time $t_c=R/c_s=180$ Myr, where $R=200$ kpc is the position of the bubble and $c_s=\sqrt{kT/(\mu m_{\rm{H}})}$ ($\mu\simeq0.62$ and $kT=5\,\rm{keV}$)~\citep{2007McNamara}. We consider that this entropy excess could also be a result of our method of using a combination of the universal pressure profile shape and the measured density profile to model the temperature profile, though we stress that this profile should have sufficient freedom to characterize the temperature at radii less than $R_{500}$. (see discussion in Section~\ref{sec::temp}).

\subsection{Luminosity and Cluster Cosmology} \label{sec::totalmass}

\begin{figure}
\begin{center}
\includegraphics[width=0.95\columnwidth]{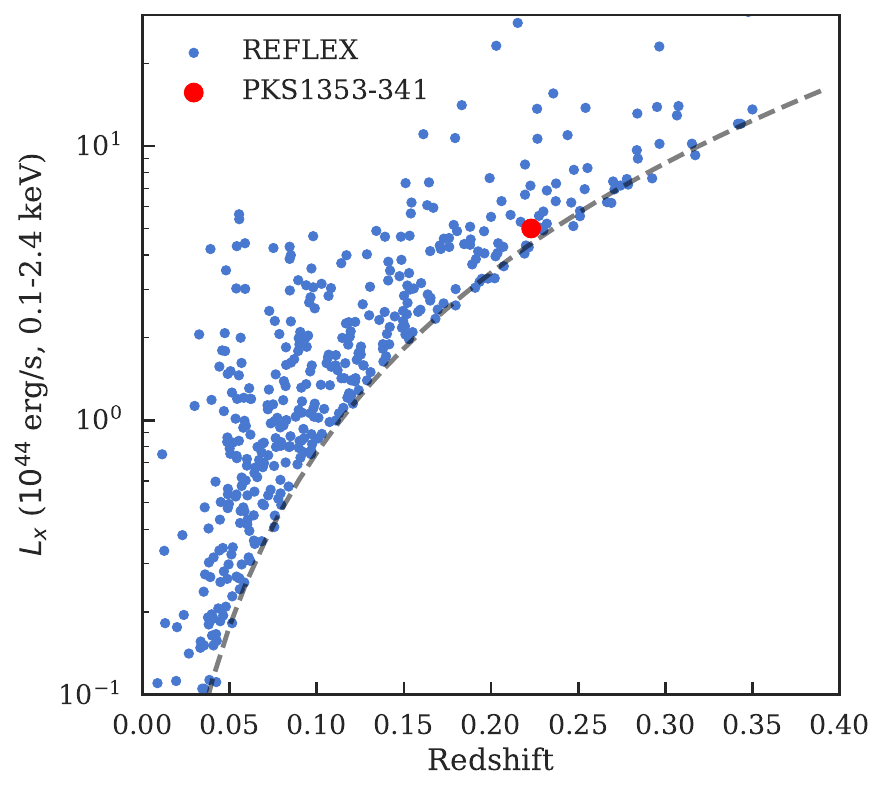}
\caption{Luminosity vs. redshift for clusters from the REFLEX Cluster Survey Catalog~\citep{2004Bohringer}, compared to PKS1353-341. The dotted line shows the cutoff introduced by~\citet{2004Bohringer} for X-ray flux limit at $3\times10^{-12}\,\rm{erg\,s^{-1}\,cm^{-2}}$. This figure demonstrates that PKS1353-341 should have been detected and identified as a cluster by REFLEX, and other cluster surveys based on \textit{ROSAT} data.}
\label{fig::luminosity_z}
\end{center}
\end{figure}

The luminosity in the \textit{ROSAT} band ($0.1\textendash2.4\,\rm{keV}$) for PKS1353-341 is $4.3\times10^{44}\,\rm{erg\,s^{-1}}$. Fig.~\ref{fig::luminosity_z} shows the luminosity of PKS1353-341 and its redshift with respect to clusters from the REFLEX all sky catalog~\citep{2004Bohringer}. The dotted line emphasizes the flux-limited nature of the REFLEX catalog at $3\times10^{-12}\,\rm{erg\,s^{-1}\,cm^{-2}}$. Comparison of the luminosity of PKS1353-341 with those from the REFLEX catalog suggests that PKS1353-341 should have been detected and identified as a cluster for its luminosity, if not for the extremely bright point source at its center.

The discovery of a new cluster above the RASS detection threshold suggests that the CHiPS survey will be able to identify similar galaxy clusters and improve upon the completeness of previous X-ray cluster surveys. Given that the constraint on $\Omega_M$ from cluster count cosmology is proportional to N$^{1/3}$~\citep{2003Haiman}, we could have a bias of $\sim$1\% in $\Omega_M$ if we are missing only a few percent of clusters due to the presence of central X-ray bright point sources. Of course, this assumes that statistics are the dominant systematic uncertainty, which is currently not the case--at present, uncertainties on $\Omega_M$ are dominated by systematic uncertainties in the cluster mass calibration. However, in the era of eROSITA (which will have a similar selection to ROSAT) and precision cluster masses, this bias may become dominant if not addressed. The CHiPS survey will provide a first estimate of how severe this bias is in the low-$z$ universe.

\subsection{Central Galaxy SFRs} \label{sec::sfr}
The SFRs of the central galaxy can be used to gauge the efficiency of cooling at the core of the cluster, assuming that the hot ICM cools and forms stars~\citep{1989McNamara,2008ODea,2011McDonald,2016McDonald}. A typical central cluster galaxy has little to no star formation; on average $\sim$1\% of the predicted cooling is observed in stars~\citep{2008ODea}, presumably because AGN feedback prevents the ICM from cooling. However, the recently discovered Phoenix cluster has a high SFR in its central galaxy, corresponding to $\sim$30\% of the predicted cooling, pointing to a different cooling mechanism or a lack of feedback at the cluster core~\citep{2012McDonald}. By computing the SFR in PKS1353-341 and comparing to the cooling rate, we can get a better understanding of the heating-cooling balance in the cluster core. 

The SFR for PKS1353-341 is computed using available archival UV data from the \textit{GALEX} Mission Archive. Based on the aperture UV flux, we estimate an SFR for PKS1353-341 of $6.2\pm3.6\,M_\odot\,\rm{yr^{-1}}$, using the empirically-derived star formation rate estimators from~\citet{2002Rosa-Gonz}, which include intrinsic extinction corrections.

\begin{figure}[!ht]
\begin{center}
\includegraphics[width=0.95\columnwidth]{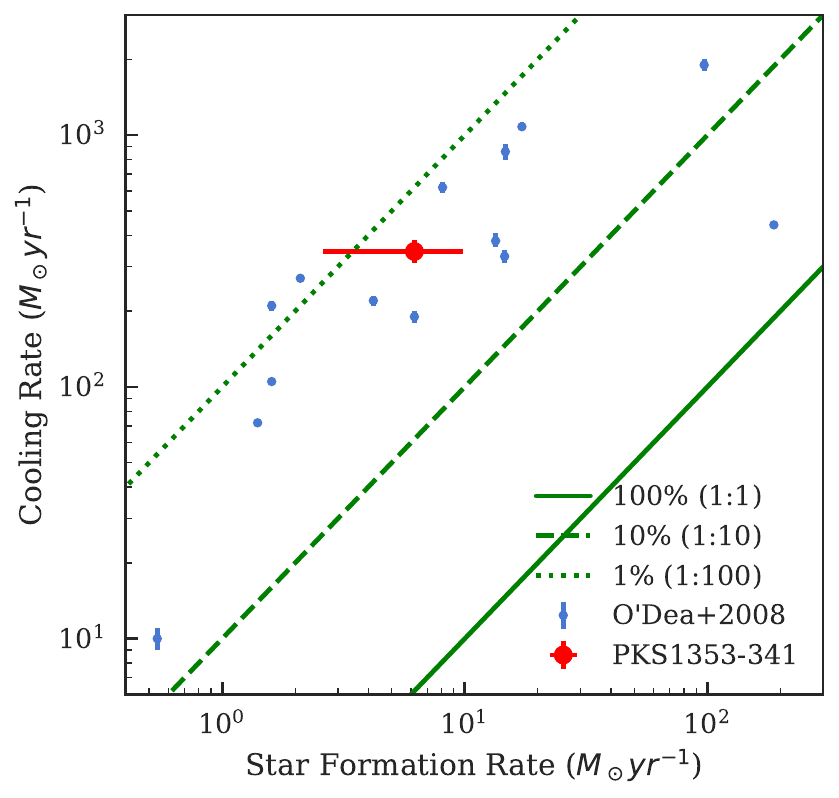}
\caption{Cooling rate vs. SFR for the central galaxy in the PKS1353-341 cluster, compared to clusters from~\citet{2008ODea}. The green lines indicate star formation proceeding at 1\%, 10\%, and 100\% efficiency with respect to the X-ray cooling rate. This figure demonstrates that cooling in PKS1353-341 is being suppressed by two orders of magnitude, presumably by AGN feedback.}
\label{fig::starformationrate}
\end{center}
\end{figure}

The SFR of $\sim$6 $M_\odot\,\rm{yr^{-1}}$ for PKS1353-341 is average, compared to those of other cool-core clusters. The SFR in this cluster should not come as a surprise since we know that cool gas is available, as evidenced by the presence of a central X-ray bright (i.e., accreting) AGN. With a larger sample size from the CHiPS survey, we could consider how the SFR depends on the two modes of AGN feedback (kinetic versus quasar). This will lead us to a better understanding of the physics around the accretion disk and how accretion relates to different feedback modes in the center of clusters. 

Nevertheless, this SFR is considered typical, compared to the ICM cooling rate, which is $345\substack{+41\\-37}\,M_\odot\,\rm{yr^{-1}}$. The SFR of PKS1353-341 is about $\sim$2\% of the cooling rate, which, according to Fig.~\ref{fig::starformationrate}, is typical for SCCs~\citep{2008ODea}. Other SCC clusters, such as A1795 and A2597, also have SFRs at this scale. Specifically, A1795's and A2597's SFR are $9$ and $4\,M_\odot\,\rm{yr^{-1}}$, respectively~\citep{2005Hicks,2007Donahue}. Thus, this system is more similar to a highly suppressed cooling flow, like A1795's, than it is to a run-away cooling flow, like the Phoenix cluster's.

\subsection{X-ray Cavities}
X-ray cavities are a result of powerful jets from central SMBHs pushing on the surrounding gas~\citep{1993Boehringer,2007McNamara}. Evidence for these cavities can be visible long after the outburst~\citep{2011Fabian}. The detection of these cavities is evidence for the existence of a radio jet, corresponding to the kinetic feedback mode of the cluster. In addition, the location and size of the cavities tell us the timescale on which these outbursts occurred and the outburst power, respectively.

\begin{figure}[!ht]
\begin{center}
\includegraphics[width=0.48\columnwidth]{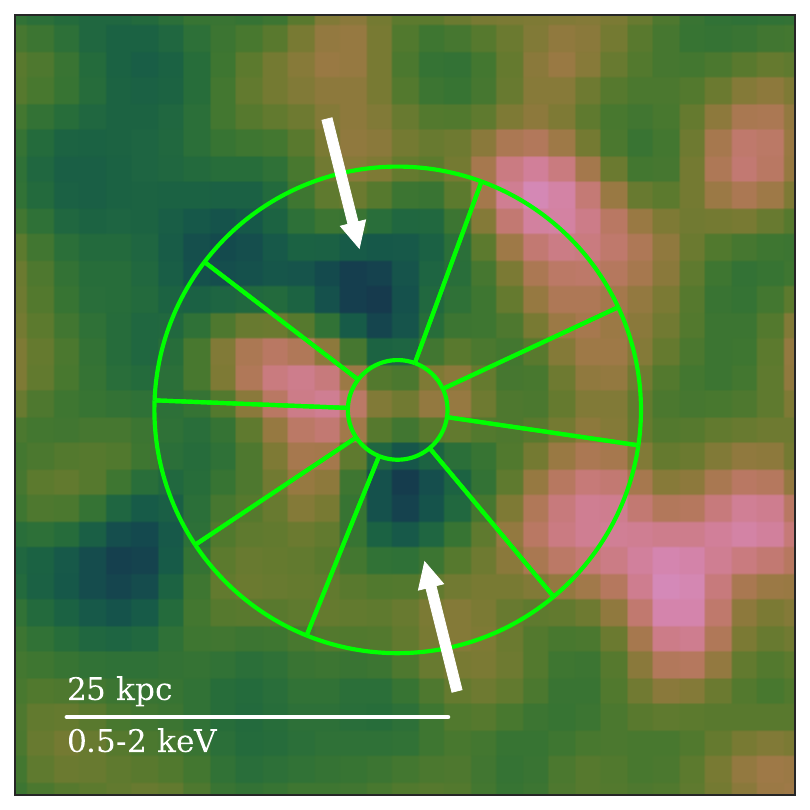}
\includegraphics[width=0.48\columnwidth]{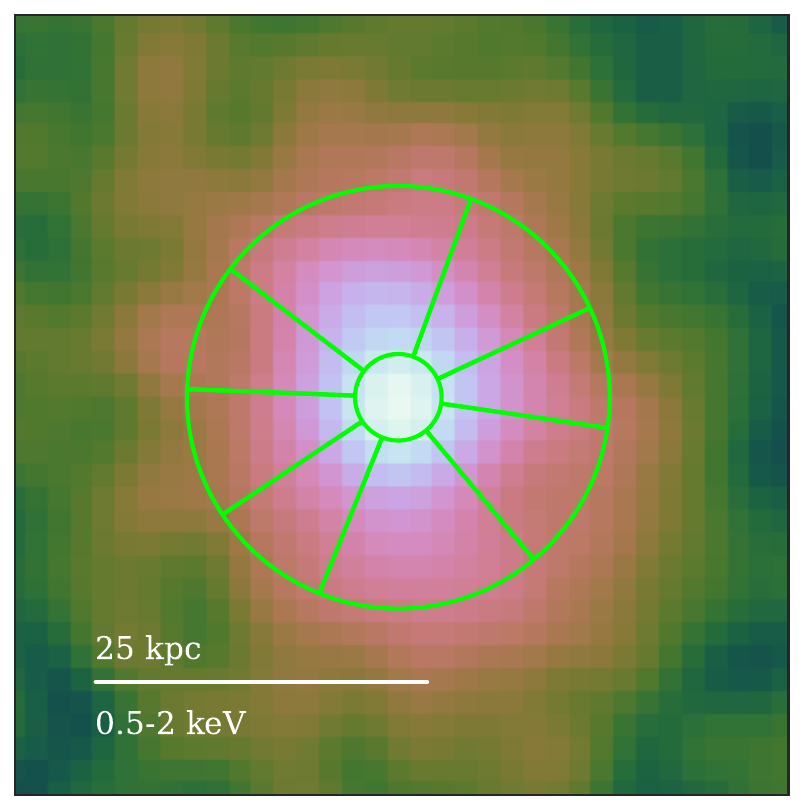}
\includegraphics[width=1.\columnwidth]{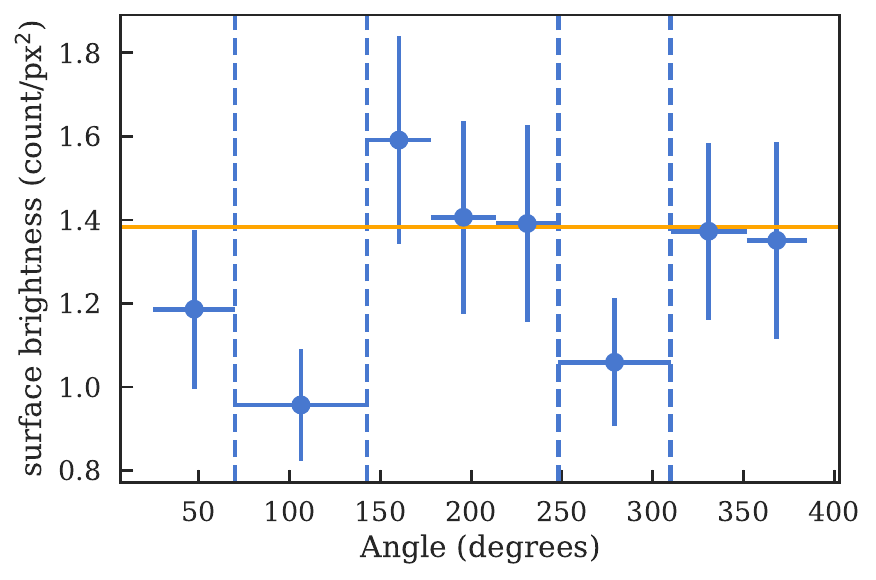}
\caption{Top left: 0.5-2 keV 2D-model-subtracted X-ray image of PKS1353-341 with potential X-ray cavities (white arrows) highlighted. Top right: 0.5-2 keV X-ray image with the regions used for calculating the azimuthal surface brightness profile in green. Bottom: 0.5-2 keV azimuthal surface brightness profile for the regions shown in the upper right panel. The vertical dashed lines show the location of each cavity (adopted from~\citet{2015Hlavacek-Larrondo}). This figure demonstrates at $\sim$2$\sigma$ depression in surface brightness at the position of the cavities.}
\label{fig::bubble}
\end{center}
\end{figure}

Using both 2D modeling and unsharp masking to remove the central bright point source and diffuse ICM, we see hints of negative residuals located symmetrically around the center point of the cluster, which may be a pair of cavities (as shown by the white arrows in the top left panel of Fig.~\ref{fig::bubble}). The top right panel of Fig.~\ref{fig::bubble} shows the annuli used for constructing the azimuthal surface brightness profile, shown in the lower panel. From this profile, we see that the potential cavities are significant at roughly the 2$\sigma$ level.

The northern cavity is located 8.5 kpc away from the center of the cluster with a cavity power $\rm{P}_{cav}=2.3\times10^{44}\,\rm{erg\,s^{-1}}$. The cavity power was calculated from the ratio between the energy stored in the bubble and the sound crossing time. The energy stored was estimated using $E_{\rm{bubble}}=4PV$ where $P$ is the thermal pressure of the ICM and $V$ is the volume of the cavity, whereas the sound crossing time was computed from $t_{\rm{cs}}=\frac{R}{c_s}$ where $R$ is the distance from the central BCG to the middle of the cavity and $c_s$ is the sound speed $\left(c_s=\sqrt{\frac{5}{3}kT/(0.62m_H)}\right)$~\citep{2015Hlavacek-Larrondo}. However, the resulting image depends strongly on the modeling and subtraction we used to remove the central point source, which led to large systematic uncertainties in the measured cavity power.

\begin{figure}[!ht]
\begin{center}
\includegraphics[width=0.95\columnwidth]{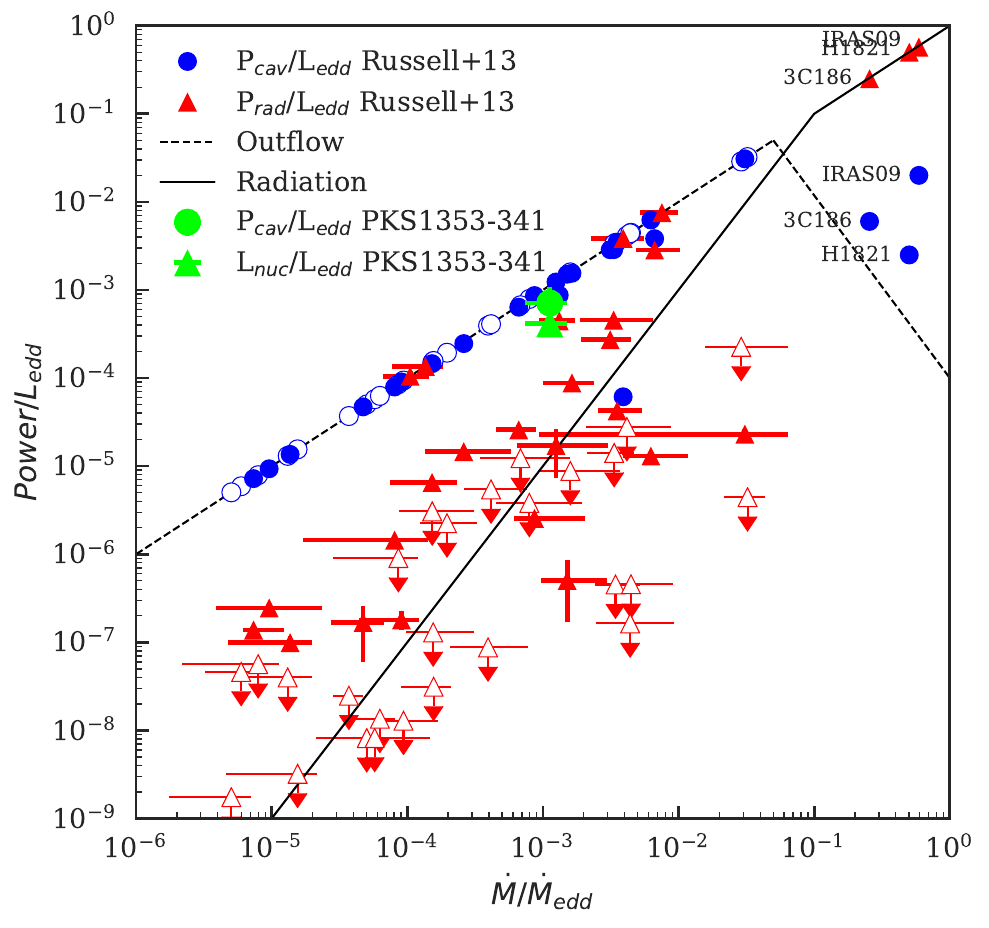}
\caption{Mean black hole accretion rates ($\dot{M}/\dot{M}_{\rm{edd}}$), compared to the cavity power ($\rm{P}_{cav}/\rm{L}_{edd}$; circle) and the radiative power ($\rm{L}_{nuc}/\rm{L}_{edd}$) of the central AGN (triangle symbols), scaled by the Eddington luminosity. Detected X-ray sources are shown as filled symbols, and the upper limits are shown in open symbols. The radiation and outflow models are for illustrative purposes (from~\citet{2013Russell}'s Fig.12). This figure demonstrates that PKS1353-341 is one of a few systems with a low accretion rate and high radiative power.}
\label{fig::cavity}
\end{center}
\end{figure}

Fig.~\ref{fig::cavity} shows the cavity power and the radiative power of PKS1353-341 with respect to the Eddington luminosity in the context of the clusters from~\citet{2013Russell}'s sample. The sample was selected from cluster, group, and elliptical galaxy samples with evidence of AGN activity in the form of cavities in X-ray images. The nuclear luminosity ($L_{nuc}\sim1.8\times10^{44}$ $\rm{erg\,s^{-1}}$) was measured from the point source at the cluster center (within 8 kpc) in the 0.1--10 keV band. For fully ionized plasma, the Eddington luminosity ($\rm{L}_{edd}$) is $1.26\times10^{47}\left(\frac{M_{\rm{BH}}}{10^9M_{\odot}}\right)\,\rm{erg\,s^{-1}}$, where $M_{\rm{BH}}$ is the SMBH mass. According to~\citet{2007Graham}, the $K$-band magnitude of the host galaxy from 2MASS can be used to estimate the SMBH mass, using $\log(M_{\rm{BH}}/M_{\odot})=-0.037(M_K+24)+8.29$, where $M_K$ is the absolute magnitude in the $K$ band. Following~\citet{2013Russell}, we assume that $\frac{\dot{M}}{\dot{M}_{\rm{Edd}}}=\frac{P_{\rm{cav}}+L_{\rm{nuc}}}{L_{\rm{Edd}}}$. 

Note that PKS1353-341, indicated as green in Fig~\ref{fig::cavity}, has $P_{\rm{cav}}$ and $L_{\rm{nuc}}$ of similar magnitude and lies on the higher end of the mean accretion rate, where the quasars (the three highest accretion rates) tend to lie. The similar magnitude of the mechanical and radiative power suggests that this system is in the midst of transitioning between mechanically dominated and radiatively dominated modes. 
\begin{figure}
\begin{center}
\includegraphics[width=0.488\columnwidth]{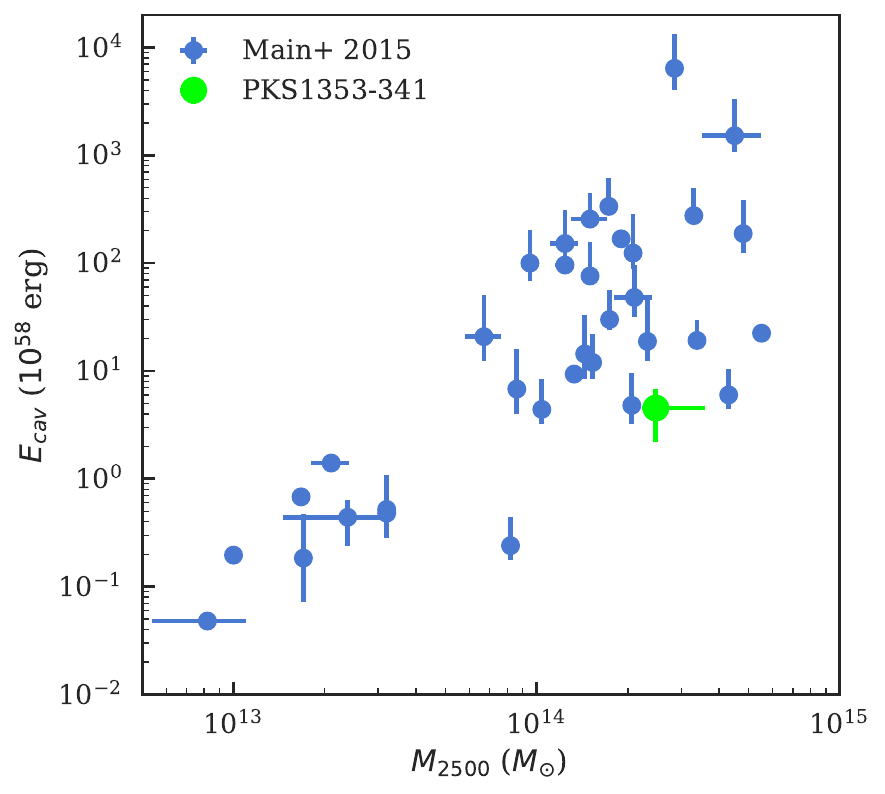}
\includegraphics[width=0.47\columnwidth]{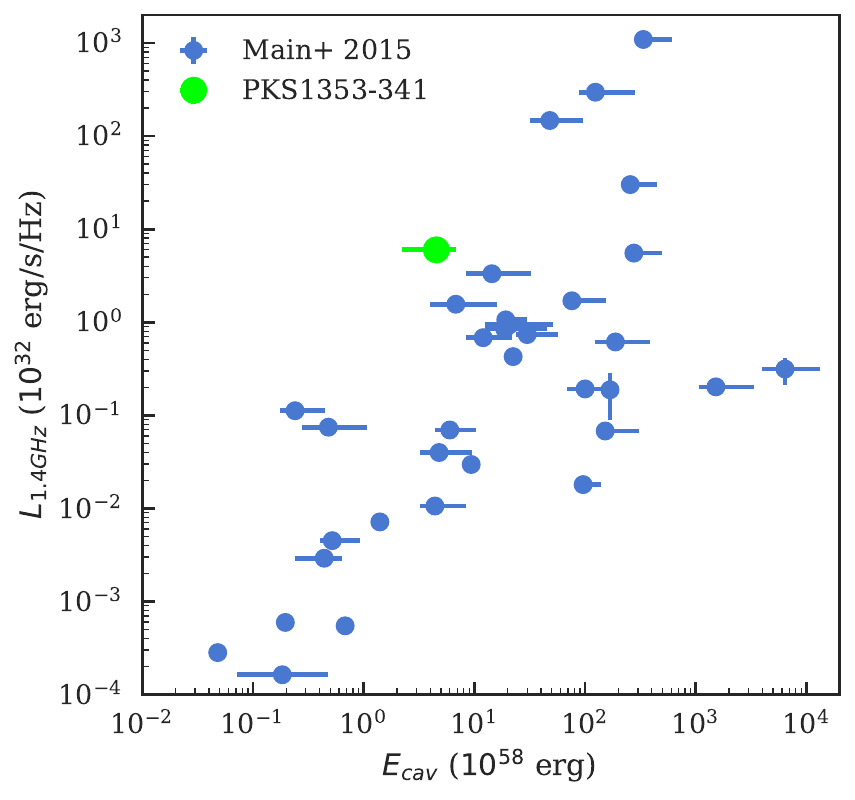}
\caption{Total energy in cavities, compared to the cluster mass (left) and the central radio power (right). PKS1353-341 is shown in green and compared to clusters from~\citet{Main2017}, shown in blue. This figure demonstrates that the amount of the mechanical feedback in PKS1353-341 is not atypical.}
\label{fig::Ecav}
\end{center}
\end{figure}

In Fig~\ref{fig::Ecav}, we plot the energy stored in the bubbles ($E_{\rm{cav}}$) for PKS1353-341 against both the total mass ($M_{2500}$) and the 1.4 GHz emission ($L_{\rm{1.4GHz}}$) compared to that of the flux-limited X-ray sample of 45 galaxy clusters from~\citet{Main2017}. The newly discovered cluster is located within the scatter in both relations, suggesting that for a cluster with this particular mass and radio power, our measured cavity energy is typical compared to those of the other clusters.

\subsection{Timescale for two different modes of AGN feedback}
In this work, we present a new galaxy cluster exhibiting possible quasar-mode feedback, which, combined with four other known QSO-hosting clusters~\citep{2010Russell,2010Siemiginowska,2012OSullivan,2012McDonald}, leads to a sample of five such systems. In comparison, the total number of known galaxy clusters exhibiting the kinetic mode is of order $\sim$100. According to the chaotic cold accretion model proposed by~\citet{2016Gaspari}, we can expect to see an increase in accretion rate of two orders of magnitude--which leads to X-ray bright nuclear sources--around 1\% of the time. This indicates that galaxy clusters with an X-ray bright nucleus should be relatively rare, only 1\% of the population. Currently, this prediction is consistent with the number of quasar-mode galaxy clusters that have been discovered, and no further mechanism is required to explain the existence of these radiatively efficient nuclei. As our survey reaches completion, we will be able to provide more evidence to either support or refute this claim.

\section{Summary}
We have presented the properties of a newly discovered galaxy cluster hosting an extreme central galaxy as the first result of a larger survey. The cluster, discovered based on its central extreme properties, is observed with the \textit{Chandra} X-ray telescope. The main results of the study are summarized as follows:

\begin{itemize}
	\item We have found a new relaxed galaxy cluster with a central X-ray and radio bright AGN in its BCG, which increases the sample size of known quasar-mode feedback clusters to about five. With the exception of the X-ray bright nucleus, this cluster is similar to other well-studied clusters such as A1795.
	\item The luminosity of the cluster is $\sim$7$\times10^{44}\,\rm{erg\,s^{-1}}$, while the average temperature is $4.3\substack{+1.7\\-1.9}\,\rm{keV}$. The total mass, assuming hydrostatic equilibrium, is $6.90\substack{+4.29\\-2.62}\times10^{14}\,M_\odot$, which makes the cluster massive enough to have been discovered by former shallow all-sky surveys (e.g., RASS). This, in combination with the discovery of the Phoenix cluster, implies that all-sky X-ray surveys can miss massive nearby cluster with bright enough central point source.
	\item The X-ray morphology, the temperature profile, the density profile, and the central cooling time all suggest that the cluster has an SCC.
	\item We find weak evidence for an excess entropy at $\sim$200 $\rm{kpc}$ from the center, which indicates a possible recent ($\sim$180 Myr) heating event that occurred near the center and is heating up and buoying high-entropy gas to a larger radius.
	\item The central galaxy has an SFR of $6.2\pm3.6\,M_\odot\,\rm{yr^{-1}}$, which is consistent with typical SFRs ($\sim1-10\,M_\odot\,\rm{yr^{-1}}$) for SCCs clusters and implies that cooling is suppressed by two orders of magnitude.
	\item We find a hint (2$\sigma$) of X-ray cavities near the cluster center. The calculated cavity power is $2.3\times10^{44}\,\rm{erg\,s^{-1}}$, which is comparable to the radiative power of the nucleus, corresponding to roughly 0.1$\%$ of the Eddington value. Deeper observations are required to confirm the existence of the cavities. This implies that the AGN may be transitioning between radiative and mechanical modes.
\end{itemize}

We have also provided an introduction to the CHiPS survey as an all-sky survey designed to find centrally concentrated galaxy cluster or clusters hosting central QSOs misidentified in previous X-ray surveys. More details about the CHiPS survey will be available in an upcoming paper.

The discovery of a new galaxy cluster implies that the CHiPS survey has the potential to increase the number of quasar-BCG clusters. This should improve our understanding of the relation between cooling and feedback processes in the cores of clusters. In addition, the survey could discover new starburst-BCG clusters, and possibly answer the uniqueness question of the Phoenix cluster. Lastly, we will be able to use the survey to help estimate the completeness of previous X-ray survey (as most current surveys are biased against clusters with central X-ray bright point sources) and give better constraints on several cosmological parameters.

\acknowledgments
Acknowledgments.\\
T.S. and M.M. acknowledge support from the Kavli Research Investment Fund at MIT, and from NASA through Chandra grant GO5-16143.


\end{document}